\documentclass[aps,preprint,nofootinbib]{revtex4}%
\pdfoutput=1
\usepackage{amsfonts}
\usepackage{amsmath}
\usepackage{amssymb}
\usepackage{float}
\usepackage{graphicx}%
\providecommand{\U}[1]{\protect\rule{.1in}{.1in}}

\begin{document}
\title{Quadratic gravity and conformally coupled scalar fields}
\author{Nicolas Caceres, Jose Figueroa, Julio Oliva, Marcelo Oyarzo, Ricardo Stuardo}
\affiliation{Departamento de F\'isica, Universidad de Concepci\'on, Casilla,
160-C, Concepci\'on, Chile.}

\email{ncaceres2016@udec.cl, josepfigueroa@udec.cl, juoliva@udec.cl, moyarzo2016@udec.cl, ricstuardo@udec.cl}

\begin{abstract}
We construct black hole solutions in four-dimensional quadratic gravity,
supported by a scalar field conformally coupled to quadratic terms in the
curvature. The conformal matter Lagrangian is constructed with powers of
traces of a conformally covariant tensor, which is defined in terms of the
metric and a scalar field, and has the symmetries of the Riemann tensor. We find exact, neutral and charged, topological black hole solutions of this theory when the Weyl squared term is absent from the action functional. Including terms beyond quadratic order on the conformally covariant tensor, allows to have asymptotically de Sitter solutions, with a potential that is bounded from below. For generic values of the couplings we also show that static black hole solutions must have a constant Ricci scalar, and provide an analysis of the possible asymptotic behavior of both, the metric as well as the scalar field in the asymptotically AdS case, when the solutions match those of general relativity in vacuum at infinity. In this frame, the spacetime fulfils standard asymptotically AdS boundary conditions, and in spite of the non-standard couplings between the curvature and the scalar field, there is a family of black hole solutions in AdS that can be interpreted as localized objects. We also provide further comments on the extension of these results to higher dimensions.

\end{abstract}
\maketitle

\section{Introduction}
Higher curvature terms in addition to the Einstein-Hilbert Lagrangian appear in many
different settings. The precise form of those correction depends on the
specific details of the theory as well as on the frame used in the
perturbative scheme that give rise to them. Higher curvature correction in four dimensions improve the renormalizability properties of the theory at the cost of spoiling its unitarity \citep{Stelle:1976gc,Stelle:1977ry}, a feature that may be absent when the whole series of corrections is considered. A particularly interesting theory with quadratic terms in the curvature in four dimensions is Critical Gravity whose Lagrangian is a precise combination of the Einstein-Hilbert term plus the Weyl square term. In this theory the coupling of the latter is fixed in terms of the negative, bared cosmological constant \citep{Lu:2011zk}. Restricting the fall-off of the asymptotically AdS
solutions in this theory, it was argued that only Einstein manifolds may
remain in the spectrum \citep{Maldacena:2011mk}. On the other hand, in asymptotically flat scenarios, the existence of black holes in General
Relativity plus the most general quadratic theory in four dimensions, leads to
spacetimes with vanishing Ricci scalar \citep{Lu:2015cqa,Lu:2015psa}.
Such argument strongly relies on the tracelessness of the Bach tensor that is
obtained as the Euler-Lagrange derivative of the conformal gravity action
$\sqrt{-g}C_{\mu\nu\lambda\rho}C^{\mu\nu\lambda\rho}$ \citep{Lu:2015cqa} ($C^\mu_{\ \nu\lambda\rho}$ being the
conformally invariant Weyl tensor). Such theory, trivially admits Einstein manifolds as solutions since the Bach tensor vanishes on spacetimes that are conformally related to the former. The spherically symmetric solutions that non-trivially depart from the Schwarzschild black hole can be integrated numerically \citep{Lu:2015cqa,Lu:2015psa}, and can also be obtained in terms of an infinite recurrence relation, which has a closed form \citep{Pravda:2016fue,Podolsky:2018pfe,Svarc:2018coe}. Such construction can be
extended to include a Maxwell field  \citep{Lin:2016kip} due to the tracelessness of its
energy-momentum tensor and a natural question therefore arises: is it
possible to consider other matter sources in this construction? Here we
explore the consequences of introducing a conformally coupled scalar field in
the action which is particularly relevant since conformally coupled scalars have been useful in the construction of black holes with secondary hair in GR \citep{BBM,Bekenstein,Martinez1,MTZ,Martinez2,Dotti:2007cp,Anabalon:2009qt,Giribet1,Giribet2,Galante1,Chernicoff:2016jsu,Chernicoff:2016uvq,Chernicoff:2016qrc}. Since the gravity Lagrangian contains operators up to mass
dimension $4$, we mimic the same structure in the matter sector, respecting a symmetry under local Weyl rescaling. A useful manner for doing so is to make use of the tensor
\begin{equation}
        S^{\mu\nu}_{\phantom{\mu\nu}\lambda\rho} 
        = \phi^{2} R^{\mu\nu}_{\phantom{\mu\nu}\lambda\rho} +\frac{4}{s}\phi\delta^{[\mu}_{[\lambda}\nabla^{\nu]}\nabla_{\rho]}\phi
        +\frac{4(1-s)}{s^{2}}\delta^{[\mu}_{[\lambda}\nabla^{\nu]}\phi\nabla_{\rho]}\phi
        -\frac{2}{s^{2}}\delta^{\mu \nu}_{\lambda \rho}\nabla_{\alpha}\phi\nabla^{\alpha} \phi \ ,\label{Stensor}
    \end{equation}
that transforms under local Weyl rescalings%
\begin{equation}
g_{\alpha\beta}\rightarrow\Omega^{2}\left(  x\right)  g_{\alpha\beta}\text{, }\phi
\rightarrow\Omega^{s}\left(  x\right)  \phi\ ,\label{transfs}
\end{equation}
as
\[
S^{\mu\nu}_{\phantom{\mu\nu}\lambda\rho}\rightarrow\Omega^{2(s-1)}(x)S^{\mu\nu}_{\phantom{\mu\nu}\lambda\rho}\ .
\]
This tensor was introduced in \citep{OlivaRay} and was used to construct conformal couplings of a scalar field to Euler densities of a higher degree,
leading to the conformally coupled version of Lovelock gravities \citep{Lovelock:1971yv}.
These couplings have allowed to construct exact black holes with secondary hair in
GR in higher dimensions \citep{Giribet1}, with interesting thermodynamic properties \citep{Giribet2,Galante1,Chernicoff:2016jsu,Hennigar:2015wxa,Hennigar:2016xwd,Hennigar:2016ekz,Dykaar:2017mba,Meng:2018wza,Mbarek:2018bau,Ghaffarnejad:2018gtj}, extending to dimension $d>4$ the known solutions in three and four dimensions of Einstein gravity and by-passing the no-go results \citep{xx,klim}.

Here we remain in dimension four, and introduce higher derivative
terms constructed out from the tensor $S^{\mu\nu}_{\phantom{\mu\nu}\lambda\rho}$. The traces of the
tensor $S$ will be important, so we define $S_{\ \beta}^{\alpha}:=S_{\ \ \beta\gamma}^{\alpha\gamma}$ and $S:=S_{\ \alpha}^{\alpha}$, which read
\begin{align}
S^{\mu\nu}&= \phi^2 R^{\mu\nu} + \frac{\phi}{s}g^{\mu\nu} \Box \phi + \frac{2\phi}{s}\nabla^{\mu} \nabla^{\nu}\phi + \frac{2(1-s)}{s^2}\nabla^\mu \phi \nabla^{\nu} \phi -\frac{(s+2)}{s^2}g^{\mu\nu} \nabla^{\rho}\phi \nabla_{\rho}\phi\ ,\label{Smunu}\\
S&= \phi^2 R +6 \frac{\phi}{s}\Box \phi -\frac{6}{s^2}\nabla^{\rho}\phi \nabla_{\rho}\phi  - \frac{6}{s} \nabla^{\rho}\phi \nabla _{\rho} \phi \ .\label{S}
\end{align}
Fixing the conformal weight of the scalar field in  \eqref{transfs} to $s=-1$, one recovers, up to a global sign, the Lagrangian for the standard conformally
coupled scalar field from the scalar quantity $S$ defined in \eqref{S}, with a canonical kinetic term \citep{improved}. In
reference \citep{OlivaRay}, the tensor $S^{\mu\nu}_{\phantom{\mu\nu}\lambda\rho}$ defined in \eqref{Stensor} was proven to be the unique, non-trivial tensorial combination of the scalar field and derivatives of the metric, that scales homogeneously under local Weyl rescalings and has the symmetries of the Riemann tensor. This tensor can also be obtained introducing a Weyl compensator for the metric which was proved useful in the construction of boundary terms in higher dimension \citep{Chernicoff:2016jsu}\footnote{At the level of the Lagrangian for the standard conformally coupled scalar field, this was done in \citep{shapiro}.}. The inclusion
of terms in the action that break the conformal invariance remove
the pure-gauge nature of the field $\phi$, making impossible to gauge it away. Similar higher curvature terms can be constructed using a Weyl-invariant curvature as explored at the quadratic level in arbitrary dimension in \cite{Dengiz:2011ig,Tanhayi:2011aa,Tanhayi:2012nn,Dengiz:2012jb}, where a gauge field $A_\mu$ on top of the metric and the scalar have to be considered. In such scenario the whole Lagrangian is Weyl invariant and therefore the scalar can be gauge fixed by a conformal transformation.

The paper is organized as follows: In Section II we present the theory consisting on the most general quadratic combination in the curvature and in the tensor $S^{\mu\nu}_{\phantom{\mu\nu}\lambda\rho}$, maintaining the conformal invariance of the matter sector. We show that due to the existence of some identities in four dimensions, the Lagrangian severely reduces. In Section III we show that the static black holes of the theory must have a constant Ricci scalar, forcing a second order constraint on the metric in a theory that has fourth-order field equations. Section IV is devoted to the construction of exact solutions describing spherically symmetric black holes in de Sitter (dS), or topological black holes with compact hyperbolic horizons in anti de Sitter (AdS). We also provide the dyonically charged version of these black holes. Section V contains a general discussion on the asymptotic behavior of the metric and the self-interacting scalar, with derivative self-interactions, admitting as a possible asymptotic behavior the standard Henneaux-Teitelboim asymptotic conditions \citep{HT}. Section VI contains an extension of these results when powers of the form $S^{k>2}$ are included maintaining the conformal invariance in the matter sector. We show that asymptotically de Sitter black holes exist even when the potential for the scalar field is bounded from below. Partial generalizations of these results to higher, even dimensions, as well as further comments are given in Section VII.

\bigskip

\section{Quadratic gravity and conformally coupled scalars}

We consider the theory%
\begin{equation}
I\left[  g_{\alpha\beta},\phi\right]  =I_{\text{g}}\left[  g_{\alpha\beta}\right]  +I_{\text{m}}\left[
g_{\alpha\beta},\phi\right]  \ ,\label{thetheory}
\end{equation}
where%
\begin{align}
I_{\text{g}}&  =\int\sqrt{-g}d^{4}x\left(  -2\Lambda
+R+\alpha_{1}R^{2}+\alpha_{2}C_{\alpha\beta\gamma\delta}C^{\alpha\beta\gamma\delta}\right.\\
&\ \ \ \ \ \ \ \ \ \ \ \ \ \ \ \ \ \ \ \ \ \ \ \ \ \ \ \ \ \ \ \ \ \ \ \ \ \ \ \ \ \ \ \ \ \ \ \ \ \ \ \ \ \ \ \ \left. +\alpha_{3}\left(  R^{2}%
-4R_{\alpha\beta}R^{\alpha\beta}+R_{\alpha\beta\gamma\delta}R^{\alpha\beta\gamma\delta}\right)  \right)  \ ,\label{Ig}\\
I_{\text{m}}&  =\int\sqrt{-g}d^{4}x\left(
-\lambda\phi^{4}-S+\beta_{1}\phi^{-4}S^{2}+\beta_{2}\phi^{-4}W_{\alpha\beta\gamma\delta}(S)%
W^{\alpha\beta\gamma\delta}(S)\right.\nonumber\\
&\ \ \ \ \ \ \ \ \ \ \ \ \ \ \ \ \ \ \ \ \ \ \ \ \ \ \ \ \ \ \ \ \ \ \ \ \ \ \ \ \ \ \ \ \ \ \ \ \ \ \ \ \ \ \ \ \left. +\beta_{3}\phi^{-4}\left(  S^{2}-4S_{\alpha\beta}S^{\alpha\beta}+S_{\alpha\beta\gamma\delta}S^{\alpha\beta\gamma\delta}%
\right)  \right)  \ . \label{Imat}%
\end{align}
where $C_{\mu \nu \rho \sigma}$ is the Weyl tensor and $W_{\mu \nu \rho \sigma}(S)$ is the traceless part of the tensor $S_{\mu \nu \rho \sigma}$, consequently defined as
\begin{align}
    W_{\mu \nu \rho \sigma}(S) = S_{\mu \nu \rho \sigma} +\left(g_{\nu[\rho}S_{\sigma]\mu}- g_{\mu[\rho}S_{\sigma]\nu}  \right)+ \frac{1}{3}S g_{\mu[\rho}g_{\sigma]\nu}\ .
\end{align}

Some remarks are now in order: in the gravitational action \eqref{Ig}, we have
explicitly considered the most general quadratic theory in a particular basis
of quadratic invariants. In four dimensions the term proportional to
$\alpha_{3}$ is the Euler density and therefore, its variation does not
contribute to the field equations. Since we are interested
only on the properties of the solutions of the theory \eqref{thetheory}, hereafter we set
$\alpha_{3}=0$. The matter action \eqref{Imat} is invariant under local Weyl rescaling,
i.e. $I_{\text{m}}\left[  \Omega^{2}\left(  x\right)  g_{\alpha\beta},\Omega^{-1}\left(
x\right)  \phi\right]  =I_{\text{m}}\left[  g_{\alpha\beta},\phi\right]  $, where we have fixed the conformal weight of the scalar field to $s=-1$, which leads to a canonical kinetic term from the scalar field which comes from the term linear in $S$. We have also
mimicked the same structure of the gravitational Lagrangian in the higher curvature terms. Since the scalar field has mass dimension $-1$, all the couplings of the
quadratic terms, $\alpha_i$ and $\beta_i$ in \eqref{Ig} and \eqref{Imat}, are dimensionless. Therefore, these are all the
terms that can appear up to quadratic order in an action, that preserve the conformal symmetry of the matter sector.

After explicitly computing the last two terms in (\ref{Imat}), one obtains%
\begin{align}
\phi^{-4}W_{\alpha\beta\gamma\delta}(S)W^{\alpha\beta\gamma\delta}(S)  &  =C_{\alpha\beta\gamma\delta}C^{\alpha\beta\gamma\delta}%
\ ,\label{LcuadradoigualaWcuadrado}\\
\phi^{-4}\left(  S^{2}-4S_{\alpha\beta}S^{\alpha\beta}+S_{\alpha\beta\gamma\delta}S^{\alpha\beta\gamma\delta}\right)   &
=R^{2}-4R_{\alpha\beta}R^{\alpha\beta}+R_{\alpha\beta\gamma\delta}R^{\alpha\beta\gamma\delta}+\text{b.t.}\ ,
\end{align}
b.t. standing for boundary terms.

The first equality (\ref{LcuadradoigualaWcuadrado}) is expected due to the
fact mentioned above regarding the tensor $S_{\alpha\beta\gamma\delta}$ as change of frame of
$R_{\alpha\beta\gamma\delta}$. Consequently, from the point of view of the field equations it is enough to consider the following action principle
\begin{equation}
I=\int\sqrt{-g}d^{4}x\left( \frac{ R-2\Lambda}{16\pi G}+\alpha
_{1}R^{2}+\alpha_{2}C_{\alpha\beta\gamma\delta}C^{\alpha\beta\gamma\delta}-\lambda\phi^{4}- \frac{1}{2}( \partial \phi)^{2} -\frac{1}{12}R\phi^{2}+\beta\phi^{-4}
S^{2}\right)  \ ,\label{TheAction}%
\end{equation}
where we redefined $\alpha_{2}\rightarrow\alpha_{2}-\beta_{2}$, and we have introduced Newton's constant. Here after we will work with this action.

\bigskip

The field equations coming from the action \eqref{TheAction}, taking variations with respect to the metric take a simple form
\begin{align}
    &\left(G_{\mu\nu} + \Lambda g_{\mu\nu}\right)+ \alpha_1\left(2g_{\mu\nu} \Box R - 2 \nabla_{\nu}\nabla_{\mu}R + 2R R_{\mu\nu} - \frac{1}{2}g_{\mu\nu}R^2\right)+ 2\alpha_2 \left(\nabla^{\alpha}\nabla^{\beta} C_{\alpha (\mu\nu)\beta} + R^{\alpha\beta} C_{\alpha(\mu\nu)\beta}\right) \nonumber\\
     +&\beta\left(\left(2 R_{\mu\nu} - 2 \nabla_{\mu}\nabla_{\nu}+g_{\mu\nu} \Box\right)(\phi^{-2}S) +12 \nabla_{(\mu}(\phi^{-3}S)\nabla_{\nu)}\phi - 6g_{\mu\nu} \nabla^{\alpha}(\phi^{-3}S\nabla_{\alpha}\phi)- \frac{1}{2}g_{\mu\nu} \phi^{-4}S^2\right)\nonumber\\
     &-\frac{1}{2}\partial_{\mu} \phi \partial_\nu \phi +\frac{1}{4}g _{\mu\nu} (\partial\phi)^2 - \frac{1}{12}\left( G_{\mu\nu} -\nabla_{\mu}\nabla_{\nu}+g_{\mu\nu} \Box \right)\phi^2+ \frac{1}{2}\lambda g_{\mu\nu}\phi^{4}=\mathcal{E}_{\mu\nu}=0 \ .\label{laecuacioncompleta}
\end{align}
The term proportional to $\alpha_2$ is the Bach tensor, while the first three terms in the third line of \eqref{laecuacioncompleta} define the standard, quadratic, improved energy-momentum tensor \citep{improved}. On the other hand, the equation for the higher curvature, conformally coupled scalar reads
\begin{align} 
\left(\Box - \frac{1}{6} R\right)\phi - 4\lambda\phi^{3} + 4\beta\left(-S^{2}\phi^{-5}+\phi^{-3}RS-3S\phi^{-4}\Box\phi -  3\Box(\phi^{-3}S)\right)=0\ .\label{ecuacionparaelescalar}%
\end{align}
As expected due to the conformal symmetry of the matter Lagrangian, the trace
of the energy-momentum that can be read from equation \eqref{laecuacioncompleta}, vanishes when one uses the equation for the scalar field (\ref{ecuacionparaelescalar}). In the following sections we will be interested in deriving some properties of the  black hole solutions of the system \eqref{laecuacioncompleta}, \eqref{ecuacionparaelescalar}. The trace of the field equations reduces to a wave equation for $R-4\Lambda$, namely
\begin{equation}
6\alpha_1\left[\Box-\frac{1}{6\alpha_1}\right]\left(R-4\Lambda\right)=0\ .\label{waveeq}
\end{equation}
As it is well known for vanishing $\Lambda$ in vacuum, equation \eqref{waveeq} after linearisation around flat spacetime leads to a massive degree of freedom of gravitational origin, with mass given by $m_{\text{eff}}^2=(6\alpha_1)^{-1}$, and therefore $\alpha_1>0$ \citep{Stelle:1976gc,Stelle:1977ry} (see also \citep{Bueno:2016ypa}). It is interesting to notice that for negative cosmological constant $\Lambda=-3/l^2$, the coupling $\alpha_1$ could be negative, provided the Breitenlohner-Freedman (BF) \citep{Breitenlohner:1982bm,Breitenlohner:1982jf} bound is fulfilled. Notice that this is not consistent with a perturbative approach since in such scenario $|\alpha_1|<<1$, and therefore the mass may be a negative large number, violating the BF bound. We therefore consider $\alpha_1>0$.

\section{Restricting the black hole solutions}
An static black hole spacetime admits the following coordinates $x^{\mu}=\left(t,x^i\right)$, where the metric takes the form
\begin{equation}
ds^2=-N^2\left(x^i\right)dt^2+h_{ij}\left(x^k\right)dx^idx^j \ .\label{staticmetric}
\end{equation}
Here $N\left(x^i\right)$ is a function that vanishes on the would-be event horizon, while $h_{ij}$ is a regular, spacelike metric in the domain of outer communications for asymptotically flat and asymptotically AdS black holes. Now, the standard argument of no-hair theorems for asymptotically flat black holes can be extended to other asymptotic behavior. Let us define the quantity $\xi=R-4\Lambda$, which by virtue of staticity depends only on the coordinates $x^i$. We will assume that the conformal scalar $\phi$ decays at infinity such that $R\rightarrow-4\Lambda$ fast enough. The wave equation in \eqref{waveeq} can be integrated on the exterior region of the black hole, which after integration by parts, leads to
\begin{equation}
\int \sqrt{h} d^3x\left(\nabla_i\left(N\xi\nabla^i\xi\right)-N\nabla_i\xi\nabla^i\xi-\frac{1}{6\alpha_1}N\xi^2\right)=0\ .\label{cuadrados}
\end{equation}
The first term in the integrand contributes as a boundary term, which acquires two contributions at the would-be horizon and at infinity. Since $N$ vanishes at the horizon, the regularity of $\xi$ implies that both contributions vanish since we have also assumed that $\xi$ approaches zero at infinity, fast enough. Since $N$ is positive in the exterior region of asymptotically flat and AdS black holes and the metric $h_{ij}$ is positive definite in that region, we find that \eqref{cuadrados} implies that $\xi$ has to vanish everywhere, and therefore, that the whole spacetime must have a constant Ricci scalar $R=4\Lambda$. While the theory is of fourth order, this restriction provides a second order constraint that helps in the explicit integration of the field equation. Notice that for asymptotically de Sitter black hole, this argument can be directly extended leading to the same conclusion that $R=4\Lambda$ in the whole region between the event and cosmological horizons. Indeed, in that case, we must integrate in the region between the horizons and since the boundary term will vanish at both horizons  we would reach the same conclusion\footnote{Notice that a similar argument has been used in \citep{Cisterna:2015iza} to restrict black hole solutions in $\mathcal{R}^2$ supergravity \citep{Ferrara:2015ela}.}. Assuming analyticity allows to extend this constraint to the whole spacetime.

\section{Exact black hole solution}
\subsection{Neutral case}
If we allow for a single metric function in a Schwarzschild like ansatz, the integration of the constraint $R=4\Lambda$ that we have derived in the previous section, fixes the metric function allowing to construct an exact solution for the whole system of equation. In fact, the metric and scalar field
\begin{equation}
ds^2=-f(r)dt^2+\frac{dr^2}{f(r)}+r^2d\Omega_\gamma^2 \ , \phi=\phi(r)\ ,\label{lineelement}
\end{equation}
define a solution of the equations \eqref{laecuacioncompleta} and \eqref{ecuacionparaelescalar}, provided $\alpha_2=0$,
\begin{equation}
f(r)=-\frac{\Lambda}{3}r^2+\gamma\left(1-\frac{\mu}{r}\right)^2\ , \ \phi(r)=\sqrt\frac{3\mu^2(1+128\pi G\Lambda(\alpha_1+\beta))}{4G\pi}\frac{1}{r-\mu}\ ,\label{solucionneutra}
\end{equation}
and
\begin{equation}
\lambda=-\frac{2\pi G\Lambda}{9(1+128\pi G\Lambda(\alpha_1+\beta))}\ .
\end{equation}
In the line element \eqref{lineelement}, $d\Omega_\gamma$ denotes the metric of an Euclidean spacetime of constant curvature $\gamma$. Notice that while the coupling of the Weyl squared term in \eqref{thetheory} has to be turned-off, this solution does receive a non-trivial contribution from the quadratic terms with couplings $\alpha_1$ and $\beta$ in \eqref{thetheory}.

The action \eqref{TheAction} contains higher curvature terms as well as conformal couplings of the scalar with quadratic curvature terms, nevertheless the form of the metric function, and therefore the causal structure of this spacetime coincides with that reported in \citep{MTZ,Martinez2}. In the spherically symmetric case $\gamma=1$ and therefore cosmic censorship implies $\Lambda>0$. Provided the parameter $\mu$ is positive and below a certain critical value $\mu_c$, the causal structure is that of Reissner-Norstrom spacetime in de Sitter space, while for $\mu=\mu_c$ the event horizon coincides with the cosmological horizon. When $d\Omega_\gamma$ is an smooth quotient of the hyperbolic space $H_2$ \citep{Mann:1996gj}, one can have event horizons even in the asymptotically AdS case, containing a rich family of causal structures that include black holes with Cauchy horizons, as well as spacetimes that can be interpreted black holes inside a black hole \citep{Martinez2}.  Notice also that the reality of the scalar field $\phi(r)$ in \eqref{solucionneutra} implies that the self-interaction potential $\lambda\phi^4$ will be bounded from below only in the asymptotically AdS case, i.e. when $\Lambda<0$. We will see below that the inclusion of extra higher curvature couplings in the matter sector, may alleviate this situation even for positive cosmological constant.

\subsection{Exact charged black hole solution}
The exact solution of the previous section can be charged, by adding to the action principle \eqref{TheAction}, the Maxwell action
\begin{equation}
I_{\text{Maxwell}}=-\frac{1}{4}F_{\mu\nu}F^{\mu\nu}\ .
\end{equation}
Assuming a dyonic ansatz for the electromagnetic field, in the Schwarzschild-like coordinates of \eqref{lineelement}, one gets
\begin{equation}
F=\frac{Q_\text{e}}{r^2}dt\wedge dr+Q_\text{m}\text{vol}(\Omega_\gamma) \ ,
\end{equation}
where $\text{vol}(\Omega_\gamma)$ represents the volume element of the two-dimensional, Euclidean manifold $\Omega_\gamma$ in \eqref{lineelement}, while $Q_\text{e}$ and $Q_\text{m}$ stand for the electric and magnetic charge, respectively. For this charged configuration, the metric function $f(r)$ and the scalar field $\phi(r)$ read
\begin{equation}
f(r)=-\frac{\Lambda}{3}r^2+\gamma\left(1-\frac{\mu}{r}\right)^2\ , \ \phi(r)=\frac{\sqrt{-6\lambda\Lambda}\mu}{6\lambda(r-\mu)}\ ,\label{solucioncargada}
\end{equation}
provided the following constraint between the integration constants $(\mu,Q_\text{e},Q_\text{m})$ is fulfilled
\begin{equation}
	Q_\text{e}^2+Q_\text{m}^2=\frac{\mu^2\gamma(9\lambda + 2\pi G \Lambda(1+576\lambda(\alpha_1+\beta)) )}{36\pi G\lambda}\ .
\end{equation}
As expected, this solution reduces to the one in the previous section when the charges $Q_{\text{e,m}}$ vanish. It is interesting to notice that this spacetime exists as a solution of the theory for arbitrary values of the couplings, and the strength of the self-interacting potential is not fixed in terms of the remaining couplings as in the uncharged case. Also notice that while the Weyl coupling $\alpha_2$ has been set to zero, the remaining higher curvature couplings $\alpha_1$ and $\beta$ in \eqref{TheAction} do contribute to the solution. Again, reality of the scalar field in \eqref{solucioncargada} imply that the cosmological constant $\Lambda$ and the algebraic, self-interacting coupling $\lambda$ must have opposite sign.

\section{On the asymptotic behavior}
The black holes constructed in the previous sections are dressed by a secondary hair, and are not continuously connected with Schwazschild-(A)dS spacetime, since as we turn off the scalar field in \eqref{solucionneutra}, the spacetime approaches the vacuum. The general solution, for arbitrary values of the coupling with primary hair has to be integrated numerically, and here we provide some general properties of such solution, when it asymptotes to a solution of GR, in vacuum. Let us focus on the asymptotically AdS case, and therefore set $\Lambda=-3/l^2$, which via the previous argument implies $R=-12/l^2$, globally. Since the contributions to the energy-momentum tensor coming from the higher curvature conformal scalar contain negative powers of the scalar, one has to check whether such terms can indeed be considered as perturbations of the standard conformally coupled scalar in the asymptotic region where the scalar field tends to zero. As a matter of fact this is the case for the particular black holes solutions we found in the previous sections, as well as in higher dimensions $d>4$ as can be seen from the explicit black hole solutions constructed in \citep{Giribet1}, but now in dimension four we have to check whether this is the case in a more generic setup. Since we are interested in the asymptotic behavior induced by the conformally coupled scalar field, we turn-off the purely gravitational higher curvature terms, i.e. we set $\alpha_1=0$ and $\alpha_2=0$ in \eqref{TheAction}. Let us introduce an effective curvature radius at infinity $l_\text{eff}$. In global AdS $f(r)=g(r)=r^2/l_\text{eff}^2+1$ and in the absence of higher curvature terms and a self-interacting potential (i.e. $\lambda=\beta=0$), the equation \eqref{ecuacionparaelescalar} reduces to that of a free, massive scalar with mass equal to the conformal mass $m^2=-2/l_\text{eff}^2$, which is above the BF mass in four dimensions $m_{\text{BF}}^2=-9/(4l_\text{eff}^2)$. For this mass, the asymptotic behavior of the scalar field, which will be consistent with the AdS asymptotics in the backreacting case, read
\begin{equation}
\phi(r)=\frac{1}{r}\left(a_0+\frac{a_1}{r}+\mathcal{O}(r^{-2})\right)+\frac{1}{r^2}\left(b_0+\frac{b_1}{r}+\mathcal{O}(r^{-2})\right)\ .\label{asympscalar}
\end{equation}
Dirichlet boundary conditions are such that $a_i=0$ while Neumann boundary conditions are defined by setting $b_i=0$. It is known that for the conformal mass, both sets of boundary conditions lead to a consistent quantization of the scalar in global AdS \citep{Breitenlohner:1982bm,Breitenlohner:1982jf}. We want to see whether this asymptotic behavior is consistent in the fully backreacting case, which is not clear at a first sight due to the presence of negative powers of the scalar field in \eqref{ecuacionparaelescalar} and \eqref{laecuacioncompleta}. We now consider in the gravitational field equations, the asymptotic expansion for the scalar field \eqref{asympscalar} as well as the following asymptotic expansion for the metric functions
\begin{equation}
f(r)=\frac{r^2}{l_\text{eff}^2}+1-\frac{m_f}{r}+\mathcal{O}(r^{-2})\ , \ g(r)=\frac{r^2}{l_\text{eff}^2}+1-\frac{m_g}{r}+\mathcal{O}(r^{-2})\ .\label{metricexpansion}
\end{equation}
Let us first analyze the case when both branches in \eqref{asympscalar} are present, i.e. $a_0,b_0\neq 0$. For the metric equations \eqref{laecuacioncompleta} one gets the following expansions
\begin{align}
\mathcal{E}^r_r&=\frac{3(l^2-l_\text{eff}^2)}{l_\text{eff}^2l^2}+\frac{3(m_f-m_g)}{r^3}+\mathcal{O}(r^{-4})\ ,\label{expansionrr0}\\
\mathcal{E}^t_t&=\frac{3(l^2-l_\text{eff}^2)}{l_\text{eff}^2l^2}+\mathcal{O}(r^{-4})=\mathcal{E}^{\theta}_{\theta}=\mathcal{E}^{\phi}_{\phi}\ .\label{expansiontt0}
\end{align}

The contribution for the matter sector \eqref{asympscalar} appears explicitly in the $\mathcal{O}(r^{-4})$ terms of \eqref{expansionrr0} and \eqref{expansiontt0}, and the vanishing of these terms, relate the arbitrary constants on the asymptotic expansions of the field and the metric functions. Notice that in all the equations, the leading terms leads to $l_\text{eff}=l$, and therefore the curvature radius at infinity is completely controlled by the bared cosmological constant and does not receive contributions from the scalar. Finally, the sub-leading term in the expansion at infinity of the field equations, relates the constants $m_f$ and $m_g$, implying therefore that the spacetimes will not only be asymptotically AdS, but asymptotically Schwazschild-AdS.

When the slow branch is not present in \eqref{asympscalar}, i.e. for Dirichlet boundary conditions for the scalar field ($a_0=0$), it is enough to consider the coefficients $b_i$ as non-vanishing. In such case, the asymptotic expansion of the components of the gravitational field equations \eqref{laecuacioncompleta} imply that $b_1$ has to vanish, and lead to
\begin{align}
\mathcal{E}^r_r&=\frac{3(l^2-l_\text{eff}^2)}{l_\text{eff}^2l^2}+\frac{72\beta(b_0l_\text{eff}^2+3b_2)}{l_\text{eff}^4b_0r^2}+\frac{3(m_f-m_g)}{r^3}+\mathcal{O}(r^{-4})\ ,\label{expansionrr}\\
\mathcal{E}^t_t&=\frac{3(l^2-l_\text{eff}^2)}{l_\text{eff}^2l^2}-\frac{72\beta(b_0l_\text{eff}^2+3b_2)}{l_\text{eff}^4b_0r^2}+\mathcal{O}(r^{-4})=\mathcal{E}^{\theta}_{\theta}=\mathcal{E}^{\phi}_{\phi}\ .\label{expansiontt}
\end{align}
This again implies that the curvature radius at infinity is determined only in terms of the bared cosmological constant. Even more, there is a single integration constant characterizing the scalar field at infinity since $b_2$ gets fixed in terms of $b_0$. The remaining orders for the equations relate the $b_{i>2}$ with $b_0$ and gravitational constant appearing in the expansion of the metric functions in \eqref{metricexpansion}. These asymptotic behavior are consistent with those of a localized matter distribution in AdS in four dimensions \citep{HT}.

Is it interesting to notice that the for the exact black holes found in the previous section, the scalar field fulfils Neumann boundary conditions, since there is a single branch at infinity, controlled by the integration constant $\mu$ that falls-off with a leading term $r^{-1}$.

\section{Beyond the quadratic level}
\begin{figure}[h!]
  \includegraphics[scale=0.4]{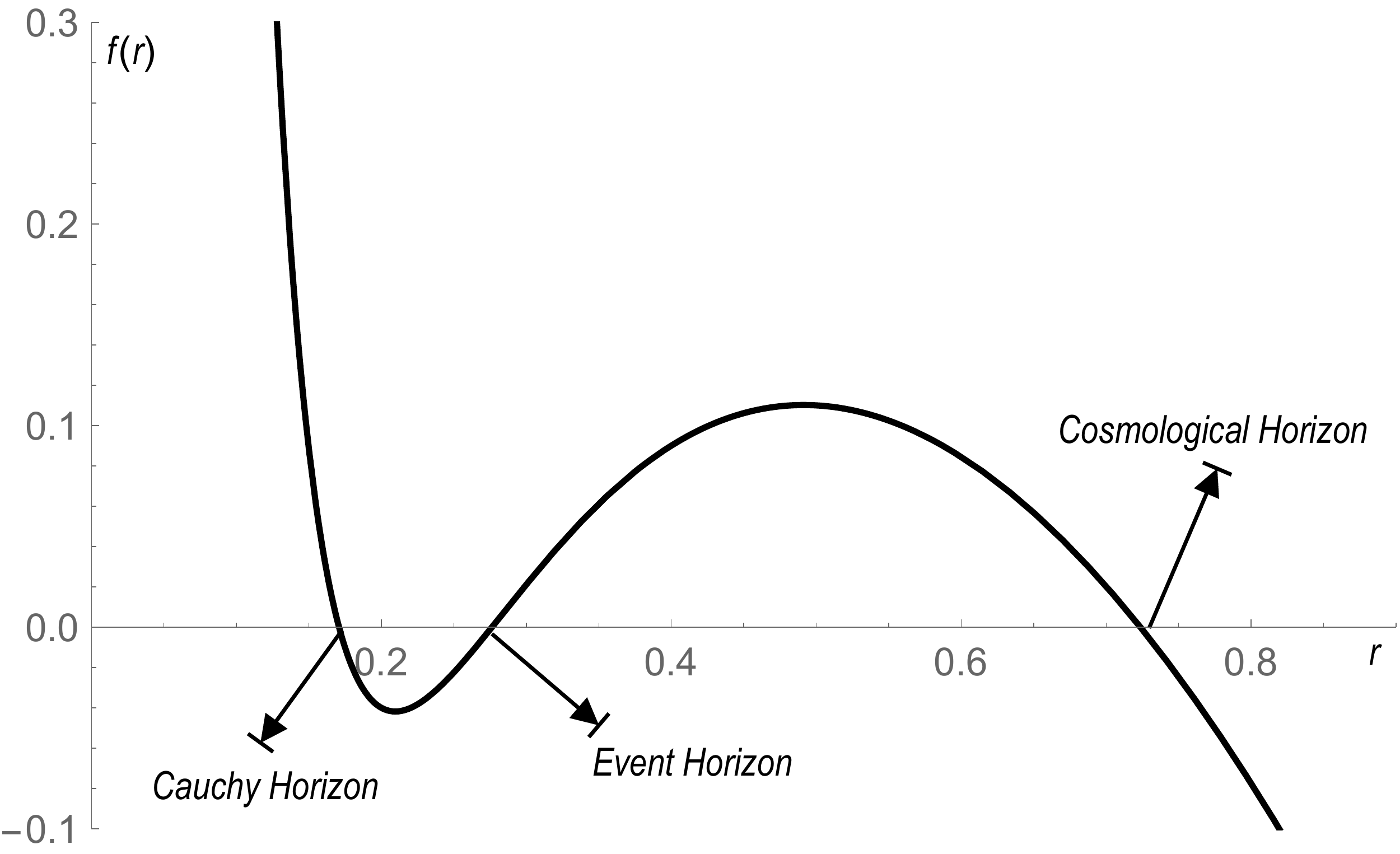}
  \caption{Four dimensional asymptotically de Sitter black hole, supported by a real, conformally coupled scalar field, with a $\lambda \phi^4$ potential that is bounded from below. The parameters defining this solution are $\Lambda=3$, $\mu=0.2$, $\beta_3=-1$ $\alpha_1=1$, $\beta=1$ and $\beta_{i>3}=0$. These parameters imply two values for the strength of the scalar field profile $a\sim.679$ and $a\sim 1.224$ in \eqref{solucionneutraalan}, but only the former leads to a potential that is bounded from below $\lambda\sim.043$ (we have also set $16\pi G=1$).}
  \label{dS}
\end{figure}

Now, lets show that the neutral solution found in section IV.A, survives the inclusion of a certain type of higher curvature, conformal couplings. Let us consider the action
\begin{equation}
I=\int\sqrt{-g}d^{4}x\left( \frac{ R-2\Lambda}{16\pi G}+\alpha
_{1}R^{2}-\lambda\phi^{4}- \frac{1}{2}( \partial \phi)^{2} -\frac{1}{12}R\phi^{2}+\beta\phi^{-4}
S^{2}+\sum_{n=3}^\infty\beta_n\phi^{4-4n}
S^{n}\right)  \ .\label{thehigheraction}%
\end{equation}
It is simple here to substitute first a metric and scalar field of the form
\begin{equation}
ds^2=-N^2(r)f(r)dt^2+\frac{dr^2}{f(r)}+r^2d\Omega_\gamma^2\ , \ \phi=\phi(r) \,
\end{equation}
in the action \eqref{thehigheraction} and then consider the field equations of the reduced action \citep{Palais:1979rca}. The solution in this case, takes the same form as before
\begin{equation}
f(r)=-\frac{\Lambda}{3}r^2+\gamma\left(1-\frac{\mu}{r}\right)^2\ , \ \phi(r)=\frac{a}{r-\mu}\ ,\label{solucionneutraalan}
\end{equation}
provided the following two constraints are fulfilled
\begin{align}
2a^2\lambda+4\Lambda\mu^2+\sum_{n=3}^\infty\beta_n (n-2)a^{-2n+2}\mu^{2n}4^{n}\Lambda^n&=0\ ,\\
\mu^2+16\pi G(8\Lambda\mu^2( \alpha_1+\beta)-a^2)+\sum_{n=3}^\infty n\beta_n\pi G\mu^{2n-2}a^{-2n+4}4^{n+1}\Lambda^{n-1}&=0 \ .
\end{align}
As before, these equations imply the existence of a single, independent integration constant that can be taken as $\mu$ or $a$. Nevertheless, the inclusion of the higher curvature terms, i.e. $\beta_{i>2}\neq 0$ allow to have spherically symmetric, asymptotically de Sitter black holes, maintaining the reality conditions of the scalar field. See Figure \ref{dS}.

\section{Further comments}
In this paper we have studied conformally coupled scalar fields containing conformal couplings with higher powers in the curvature, in four dimensions. We have shown that the solution found in \citep{MTZ} can be embedded in these theories, provided the relations between the couplings are updated due to the presence of the new terms in the action. In contrast with what occurs in such reference, the presence of the new terms allow to construct asymptotically de Sitter black holes with self-interactions of the form $\lambda \phi^4$ which are bounded from below. These solutions fulfil Neumann boundary conditions, which are still consistent in the backreacting case, even in the presence of the non-standard derivative, conformal, self-interactions of the scalar field.

We proved that even in the presence of quadratic terms in the curvature in the purely gravitational action, as well as a cosmological term, one has that the static, regular black holes must have a constant Ricci scalar. Such argument can be trivially extended to a certain family of theories in higher dimensions as follows. For simplicity we focus on the gravity part, but the argument extends to conformally invariant matter sources. Consider the following theory in dimension $D=4k$%
\begin{equation}
I_{4k}=\int\sqrt{-g}d^{D}x\left(  R+\alpha_1 R^{2k}+\alpha_2\ tr\left(
C^{2k}\right)  \right)  \ ,\label{I4k}
\end{equation}
where $\alpha_{1,2}$ are are arbitrary, dimensionless, coupling constants.
The field equations are of fourth order and or the form%
\begin{equation}
G_{\mu\nu}+\alpha_2 B_{\mu\nu}+\alpha_1 E_{\mu\nu}=0\ ,\label{ecusarbd}
\end{equation}
where $B_{\mu\nu}$ is a  traceless, generalized Bach tensor, that comes from the variation of the Weyl invariant Lagrangian $\ tr\left(
C^{2k}\right)$
and%
\begin{equation}
E_{\mu\nu}=2kR^{2k-1}\left(  R_{\mu\nu}-\frac{1}{4k}g_{\mu\nu}R\right)
+2kg_{\mu\nu}\square R^{2k-1}-2k\nabla_{\mu}\nabla_{\nu}R^{2k-1}\ .
\end{equation}

The trace of the field equations \eqref{ecusarbd} reads%
\begin{equation}
\left(  1-2k\right)  R+\left(  8k^{2}-2k\right)  \alpha_1 \square R^{2k-1}=0\ .
\end{equation}
Assuming the general ansatz for a static spacetime given in \eqref{staticmetric}, leads to%
\begin{align}
\square R^{2k-1} &  =\frac{1}{N\sqrt{h}}\partial_{i}\left(  N
\sqrt{h}h^{ij}\partial_{j}R^{2k-1}\right)=\frac{\nabla_{i}N}{N}\nabla^{i}R^{2k-1}+\nabla^{2}R^{2k-1}\ .
\end{align}
Where $\nabla_{i}$ is the covariant derivative constructed with the Levi-Civita
connection associated to $h_{ij}$. Then, after multiplying the trace of the
field equations by $R^{2k-1}N$ one gets%
\begin{align}
\left(  1-2k\right)  R^{2k}N+2k\left(  4k-1\right)  \alpha_1\left(
\nabla_{i}\left(  N R^{2k-1}\nabla^{i}R^{2k-1}\right)  -N \nabla_{i}R^{2k-1}%
\nabla^{i}R^{2k-1}\right)   &  =0\ ,
\end{align}
and then we integrating on the spacelike section and assuming that the boundary term vanishes as in Section III, we get
\begin{equation}
\int\sqrt{h}d^{D-1}x\left(  \left(  1-2k\right)  R^{2k}N-2k\left(
4k-1\right)  \alpha_1 N \nabla_{i}R^{2k-1}\nabla^{i}R^{2k-1}\right)  =0\ .
\end{equation}
Then, assuming $\alpha_1>0$ we have that the vanishing of the integral, requires the vanishing of each of the terms in the integrand, therefore%
\begin{equation}
R=0\ ,
\end{equation}
generalizing the results of the previous sections to dimensions $D=4k$. Notice that in the action \eqref{I4k}, we have considered a particular combination of Weyl invariant terms. In dimension four, there exists a single term with this property, namely $C^2$, while in dimension eight, there are seven, algebraic invariants that are constructed by complete contractions of the four Weyl tensors. At the level of spherically symmetric spacetimes or more generally spacetimes of the form \eqref{lineelement}, all these terms contribute with the same functional form to the field equations \citep{Deser:2005pc}. 

It is well know that solutions of GR with a standard conformally coupled scalar field can be embedded in eleven-dimensional supergravity (see e.g. \citep{deHaro:2006ymc}). It would be interesting to see whether the corrections to the conformal coupling we have introduced here can be mapped with some corrections of the eleven-dimensional theory.

\section*{Acknowledgements}
We thank Andrés Anabalon, Adolfo Cisterna and Gaston Giribet for useful comments. The research of J.O. is supported in part by the Fondecyt Grant 1181047. J.F. and R.S. thank the support of CONICYT through the Fellowships 22191705 and 22191591, respectively.

\end{document}